\documentclass[prb,twocolumn,showpacs,superscriptaddress]{revtex4}

\usepackage{graphicx}
\usepackage[centertags]{amsmath}

\newcommand{\cm}{\ensuremath{\,\mbox{cm}^{-1}}}
\newcommand{\K}{\ensuremath{\,\mbox{K}}}
\newcommand{\celsius}{\ensuremath{\,{}^\circ}\!C}

\begin{document}

\title{Dielectric relaxation and polar phonon softening in relaxor ferroelectric PbMg$_{1/3}$Ta$_{2/3}$O$_3$ }

\author{ S.~Kamba}\email{kamba@fzu.cz}
\affiliation{Institute of Physics ASCR, v.v.i. Na Slovance~2, 182 21 Prague~8, Czech
Republic}
\author{D.~Nuzhnyy}
\affiliation{Institute of Physics ASCR, v.v.i. Na Slovance~2, 182 21 Prague~8, Czech
Republic}
\author{S.~Veljko}
\affiliation{Institute of Physics ASCR, v.v.i. Na Slovance~2, 182 21 Prague~8, Czech
Republic}
\author{V.~Bovtun}
\affiliation{Institute of Physics ASCR, v.v.i. Na Slovance~2, 182 21 Prague~8, Czech
Republic}
\author{J.~Petzelt}
\affiliation{Institute of Physics ASCR, v.v.i. Na Slovance~2, 182 21 Prague~8, Czech
Republic}
\author{Y.L. Wang}
\affiliation{Ceramics Laboratory, Swiss Federal Institute of Technology - EPFL CH-1015
Lausanne, Switzerland}
\author{J. Levoska}
\affiliation{Microelectronics and Materials Physics Laboratories, University of Oulu, PL
4500, FIN 90014 Oulun Yliopisto, Finland}
\author{M. Tyunina} \affiliation{Microelectronics and Materials Physics
Laboratories, University of Oulu, PL 4500, FIN 90014 Oulun Yliopisto, Finland}
\author{J. Macutkevic}
\affiliation{Semiconductor Physics Institute, A. Gostauto 11, LT-2600 Vilnius, Lithuania}
\author{N. Setter}
\affiliation{Ceramics Laboratory, Swiss Federal Institute of Technology - EPFL CH-1015
Lausanne, Switzerland}
\author{J. Banys}
\affiliation{Faculty of Physics, Vilnius University, Sauletekio 9, LT-10222 Vilnius ,
Lithuania}
\date{\today}


\begin{abstract}

Relaxor ferroelectric PbMg$_{1/3}$Ta$_{2/3}$O$_3$ ceramics and thin films were
investigated by means of broad-band dielectric, time-domain terahertz (THz) and
Fourier-transform infrared (IR) spectroscopy in the frequency range 100 Hz - 90 THz at
temperatures 100 - 490\K, the THz and IR spectra were studied from 20 to 900\K. Diffused
and strongly temperature dependent peak in the complex permittivity is caused by a
dielectric relaxation due to the dynamics of polar clusters. The relaxation appears below
Burns temperature $T_{d}$ in the THz range, slows down on cooling through the microwave
and MHz range and anomalously broadens. The shortest and longest relaxation times of the
distribution of relaxation times follow Arrhenius and Vogel-Fulcher law, respectively.
The degree of B-site order has only a small influence on the parameters of the dielectric
relaxation and almost no influence on the phonon parameters. Below T$_{m}\cong$180\K\,
the distribution of relaxation frequencies becomes broader than our experimental spectral
range and frequency independent dielectric losses develop below 100 GHz in the spectra.
Although the macroscopic crystal structure is cubic, IR spectra give evidence about the
lower local symmetry which can be assigned to the presence of polar clusters below
$T_{d}$. Infrared spectra above $T_{d}$ still reveal more modes than predicted by
selection rules in the paraelectric phase of the $Fm\bar{3}m$ space group so that we
suggest selection rules which take into account chemical inhomogeneity in the
$\beta$''-perovskite sublattice.

\end{abstract}

\maketitle

\section{Introduction}
The relaxor ferroelectricity in complex perovskites like
PbMg$_{1/3}$Nb$_{2/3}$O$_3$ (PMN) has been intensively
investigated since the early work by Smolenskii and Agranovskaya
\cite{smolenskii58} and the interest in this class of dielectrics
underwent a revival after the pioneering work by Park and Shrout
\cite{park97}, who discovered ultrahigh strain and giant
piezoelectric response in PbMg$_{1/3}$Nb$_{2/3}$O$_3$-PbTiO$_{3}$
and PbZn$_{1/3}$Nb$_{2/3}$O$_3$-PbTiO$_{3}$ single crystals. The
relaxor ferroelectrics exhibit a high and broad peak in the
temperature dependent permittivity whose position remarkably
shifts to higher temperatures with increasing measuring frequency.
The peculiar dielectric properties are caused by a wide dielectric
relaxation which broadens and slows down on
cooling.\cite{viehland} The relaxation originates from the
dynamics of polar clusters (1-6\,nm in diameter), which develop
below the Burns temperature T$_{d}$ (600-700\,K in various complex
perovskites)\cite{burns83} and it is believed that the broad
distribution of relaxation frequencies has its origin in random
fields and random forces as a consequence of chemical disorder in
the perovskite B sites occupied by two cations of different
valence (e.g. Mg$^{2+}$ and Nb$^{5+}$ in PMN).\cite{westphal92}
Therefore it appears that the degree of chemical order of B-site
cations plays a crucial role in the dielectric response of
perovskite relaxors.

It was shown that the B cations in
Pb(B'$_{1/3}^{2+}$B''$_{2/3}^{5+}$)O$_{3}$ (B'=Mg, Ni, Zn and
B''=Nb, Ta) perovskites reveal the 1:1 order in small domains
(2-9\,nm) dispersed in a disordered perovskite matrix. In the 1:1
ordered domains with general formula
Pb($\beta'_{1/2}\beta''_{1/2}$)O$_3$ the $\beta'$ sub-lattice is
occupied solely by B'' cations, but the $\beta''$ one contains a
random distribution of B' and of the remaining B'' cations in the
ratio 2:1.\cite{akbas97,yan98,xu00} In this description the
ordered structure of PMN, for example, can be represented as
Pb(Mg$_{2/3}$Nb$_{1/3}$)$_{1/2}$Nb$_{1/2}$O$_3$.\cite{yan98,xu00,davies00}
First-principles-based simulations have shown that the polar
nanoregions appear predominantly in chemically ordered
regions.\cite{burton06} The size of 1:1 ordered domains can be
enlarged by suitable doping or sometimes by longer annealing the
ceramics. For example, in PMN the ordered domains were enlarged by
two orders of magnitude (up to 500\,nm) by 30\% La doping and
simultaneously the permittivity was remarkably reduced to 200
(from $\sim$20000 in pure PMN), but a weak dielectric relaxation
still remained present.\cite{chen89}

In closely related PbMg$_{1/3}$Ta$_{2/3}$O$_3$ (PMT) ceramics the
1:1 ordering can be achieved by proper high-temperature heat
treatment, but the highly ordered samples still contain a large
volume of disordered anti-phase boundaries and the 1:1 domains do
not coarsen over $\sim$10\,nm. More pronounced domain growth (up
to $\sim$40\,nm) and up to 95 vol. \% ordering has been observed
in PMT that contain small admixtures of
PbZrO$_{3}$.\cite{akbas97,akbas01,dmowski02} Nevertheless, in
spite of the domain growth, only small influence of the ordering
degree on the dielectric relaxation was observed - magnitude and
temperature of the maximum permittivity ($\varepsilon'_{max}$ and
$T_{m}$) slightly reduce with
ordering.\cite{akbas97,akbas01,wang05} It therefore appears that
the randomness of the ordered structure in the $\beta''$
sublattice is the critical factor for appearance of the relaxor
ferroelectric behavior which inhibits the normal ferroelectric
coupling\cite{akbas01} rather than the chemical cluster size.

Numerous publications exist about the dielectric properties of
PMT,\cite{akbas97,akbas01,dmowski02,wang05,bokov61,lu96,ko03} however, all the
measurements were performed only up to 1\,MHz and mostly below room temperature. For
understanding the relaxation behavior, higher-frequency dielectric spectra at
temperatures above room temperature, especially near and above the Burns temperature
$T_{d}$, are necessary. We will report here about dielectric spectra measurements between
100\,Hz and 9\,GHz at temperatures up to 490\K\, and about the dielectric response in the
THz and infrared (IR) spectral range (0.1 - 90\,THz) for 20 - 900\K. Particularly the THz
and IR spectra near $T_{d}$ are of importance since in several other relaxors we observed
that the dielectric relaxation moves from radio-frequency range near $T_{m}$ up to the
THz range near $T_{d}$ and merges with polar phonons above
$T_{d}$.\cite{kamba05,kamba05b,macutkevic06,bovtun06} Moreover, the local ferroelectric
instability in the polar clusters was indicated in the IR spectra by an unstable polar
optic phonon which partially softens on heating to $T_{d}$ and hardens above
it.\cite{wakimoto02b,kamba05,kamba05b,macutkevic06,bovtun06} The question to what extend
this behavior is general needs still to be investigated not only near and above $T_{d}$,
but also below $T_{m}$. Simultaneously, it will be of interest to study the influence of
the 1:1 order degree on the IR spectra especially above $T_{d}$. For this purpose we
processed and investigated three types of PMT ceramics of the order from 10 to 90\%. It
is worth to note that except for the very recent paper on PLZT\cite{buixaderas07}, above
$T_{d}$ no IR reflectivity spectra of relaxors were reported up to now.

\section{Experimental}

To achieve a different degree of the 1:1 order in PMT ceramics we added 5\% of
PbZrO$_{3}$ into PMT. Details of the ceramics processing were described in
Ref.\cite{wang05}. The as-sintered ceramics (at 1230\celsius\, for 2 hours) have less
than 10\% order (domain size $\sim$3-5\,nm) while the ceramics annealed for 1 hour at
1350\celsius\, exhibit 15\% ordering (domain size $\sim$3-5\,nm) and the 64-hours
annealed ceramics show 90\% order and ordered domain size of about 150\,nm. The phase
content in ceramics was monitored by a powder XRD diffractometer Siemens Kristalloflex
805 (Cu K$\alpha$, 40 kV, 35 mA) and conventional TEM (Philips CM300-UT-FE). Details of
the order degree determination in PMT ceramics (from the intensity of the
$\frac{1}{2}$(111) superlattice reflections) was published in Ref.\cite{wang05}

PMT thin film (of 490\,nm thickness) was grown on a (0001) oriented sapphire substrate
with 40\,nm thick SrTiO$_{3}$ (STO) buffer layer using $\textit{in situ}$ pulsed laser
deposition\cite{tyunina04} at the substrate temperature of 650\celsius. The
room-temperature x-ray (Cu K ) diffraction studies showed that the film was highly (111)
oriented polycrystalline. No indication of B-site order was observed by XRD.

PMT single crystal was grown by the flux method described in Ref.
\cite{kania06} (kindly provided by A. Kania from the University of
Silesia, Katowice, Poland). The crystal size was 5x5x4 mm. The
crystal had many cracks inside so that it was difficult to polish
it for IR and THz measurements. Due to imperfect surface of the
cut crystal plates, the crystal was used just for qualitative
comparison with ceramics and thin films.

The low-frequency dielectric properties were measured as a
function of temperature in the 100 Hz - 1 MHz frequency range
using a high precision LCR meter (HP4284A) and a
computer-controlled environmental chamber (Delta 9023).
High-frequency response (1\,MHz - 1.8\,GHz) was obtained by means
of the impedance analyzer Agilent 4291B with a Novocontrol BDS
2100 coaxial sample cell and a Sigma System M18 temperature
chamber (operating range 100 - 570\K). The TE$_{0n1 }$ composite
dielectric resonator method\cite{krupka06} and network analyzer
Agilent E8364B were used for microwave measurements at 8.8 GHz in
the 100 - 350\K\, temperature interval. The cooling rate was
2\K/min.

Measurements at THz frequencies from 7 to 33 cm$^{-1}$ (0.2 -
1.0\,THz) were performed in the transmission mode using a
time-domain THz spectrometer based on an amplified Ti - sapphire
femtosecond laser system. Two ZnTe crystal plates were used to
generate (by optic rectification) and to detect (by electro-optic
sampling) the THz pulses. Both the transmitted field amplitude and
phase shift were simultaneously measured; this allows us to
determine directly the complex dielectric response
$\varepsilon^{\ast}(\omega)$. An Optistat CF cryostat with mylar
windows (Oxford Inst.) was used for measurements down to 20\K.

IR reflectivity and transmission spectra were obtained using a
Fourier transform IR spectrometer Bruker IFS 113v in the frequency
range of 20 - 3000 cm$^{-1}$ (0.6 - 90 THz) at room temperature.
At lower temperatures the reduced spectral range up to 650
cm$^{-1}$ only was studied (transparency region of the
polyethylene windows in the Optistat CF cryostat). Pyroelectric
deuterated triglycine sulfate detectors were used for the room-
and high-temperature reflectivity measurements, while more
sensitive liquid-He cooled (1.5 K) Si bolometer was used for the
low-temperature measurements as well as for high-temperature IR
transmission experiment. Polished disk-shaped ceramic samples with
a diameter of 8 mm and thickness of $\sim$1 mm were used for IR
reflectivity studies, while the thin film on sapphire substrate
(thickness 05\,mm) was investigated in the transmission mode.

\section{Results and discussions}

\begin{figure}
  \begin{center}
    \includegraphics[width=80mm]{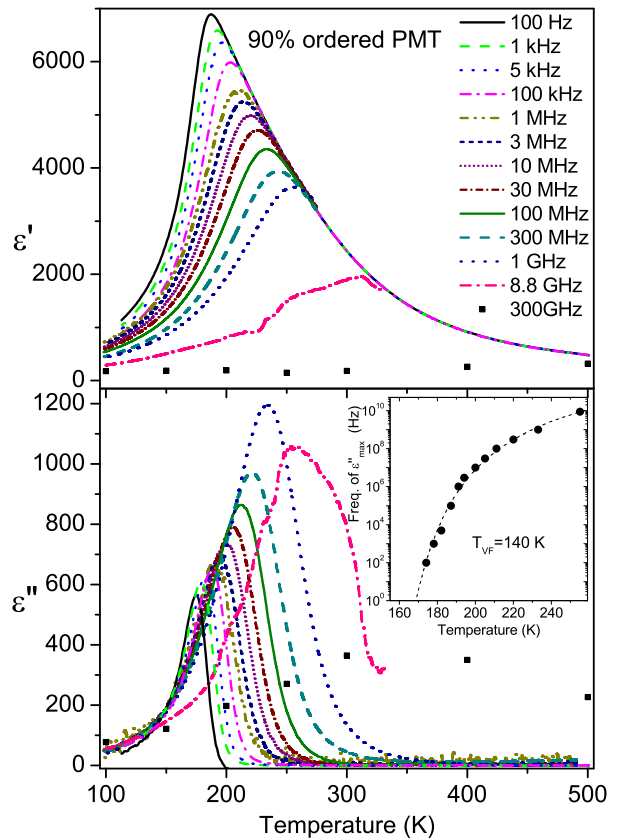}
  \end{center}
    \caption{(color online) Temperature dependence of the real $\varepsilon$' and imaginary $\varepsilon$'' part of complex permittivity
    in 90\% ordered PMT ceramics at different frequencies between 100\,Hz and 300\,GHz. Inset shows the Vogel-Fulcher fit of
    $\varepsilon''_{max}$(T) frequencies yielding the freezing temperature $T_{VF}$=140\K.}
    \label{Fig1}
\end{figure}

\begin{figure}
  \begin{center}
    \includegraphics[width=85mm]{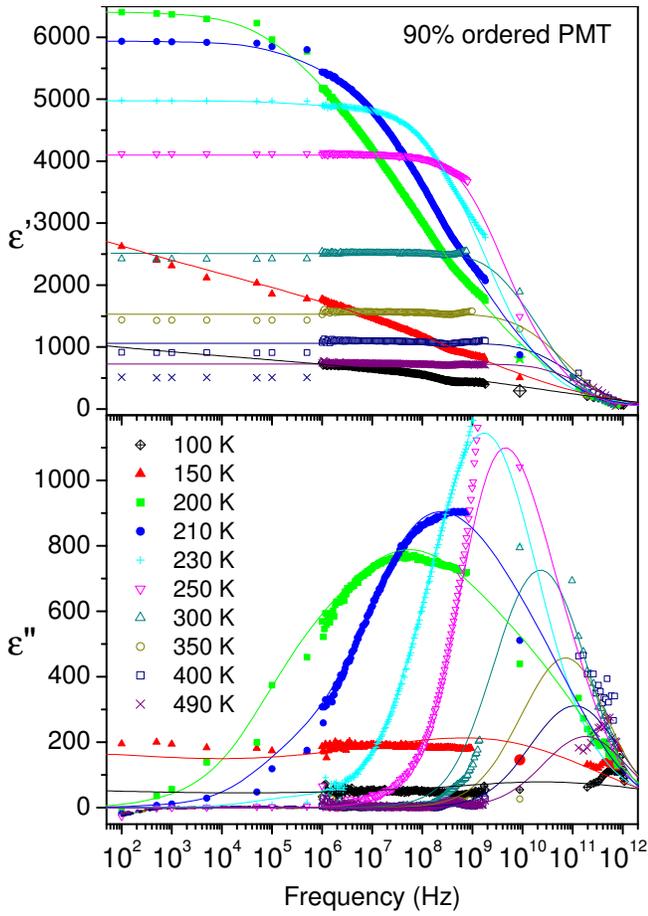}
  \end{center}
    \caption{(color online) Frequency dependence of the complex permittivity in 90\% ordered PMT ceramics at various temperatures.
    Points are experimental data, the lines are results of the fit with Eq.~\ref{integr}. }
    \label{Fig2}
\end{figure}

Temperature dependence of the real and imaginary part of complex
permittivity $\varepsilon^{\ast}= \varepsilon^\prime - {\rm
i}\varepsilon^{\prime\prime}$ for the 90\% ordered PMT ceramics at
various frequencies is plotted in Fig.~\ref{Fig1}. Less ordered
PMT ceramics exhibit qualitatively similar dielectric anomaly with
$T_{m}$ and $\varepsilon'_{max}$ shifting to higher values (about
30\% and 5\% shifts up of $\varepsilon'_{max}$ and $T_{m}$,
between PMT ceramics of 90\% and 10\% order,
respectively).\cite{wang05} Therefore we do not present here the
data of all variously ordered samples. Frequency dependence of the
complex permittivity between 100\,Hz and 1\,THz is plotted in
Fig.~\ref{Fig2}. It shows a broad dielectric relaxation which
cannot be fitted with a simple Debye formula. Therefore a more
general model of distribution of relaxation times $f$($\tau$) is
used for modeling the complex permittivity. In this case the
$\varepsilon^\prime$ and $\varepsilon^{\prime\prime}$ are
expressed as
\begin{subequations}\label{integr}
\begin{eqnarray}
\label{integ1}
    \varepsilon(\omega)'=\varepsilon_{R\infty}+\Delta\varepsilon \int_{-\infty}^{\infty}\frac{f(\tau)d(ln\tau)}{1+\omega^2\tau^{2}},\\
\label{integ2}
    \varepsilon"(\omega)=\Delta\varepsilon \int_{-\infty}^{\infty}\frac{\omega\tau f(\tau)d(ln\tau)}{1+\omega^2\tau^{2}}.
\end{eqnarray}
\end{subequations}
with the normalization condition
\begin{equation}\label{normalization}
\int_{-\infty}^{\infty}f(\tau)d(ln\tau)=1.
\end{equation}

\begin{figure}
  \begin{center}
    \includegraphics[width=85mm]{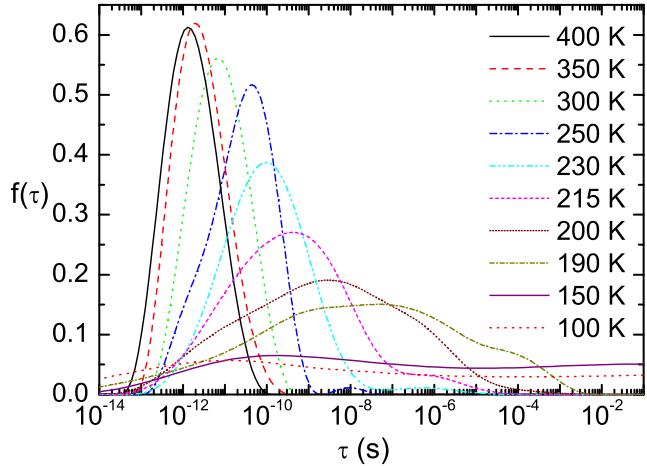}
  \end{center}
    \caption{(color online) Distribution function of relaxation times in the 90\% ordered PMT ceramics calculated at various
    temperatures from the fits of complex permittivity by means
    of Eq.~\ref{integr}. Note that the relaxation frequency is $f_{r}$=2$\pi/\tau$.}
    \label{Fig3}
\end{figure}

\begin{figure}
  \begin{center}
    \includegraphics[width=85mm]{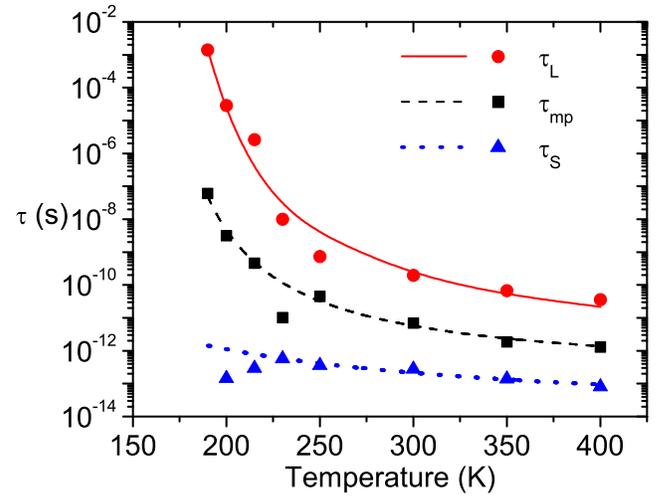}
  \end{center}
    \caption{(color online) Temperature dependence of the longest ($\tau_{L}$), most probable ($\tau_{mp}$) and shortest ($\tau_{S}$)
    relaxation times in the 90\% ordered PMT ceramics. $\tau_{mp}$ corresponds to maximum of $f(\tau)$. $\tau_{L}$ and $\tau_{S}$ are defined
    as edges of the distribution function where $f(\tau)$=0.1$f(\tau_{mp})$. }
    \label{Fig4}
\end{figure}

\begin{figure}
  \begin{center}
    \includegraphics[width=85mm]{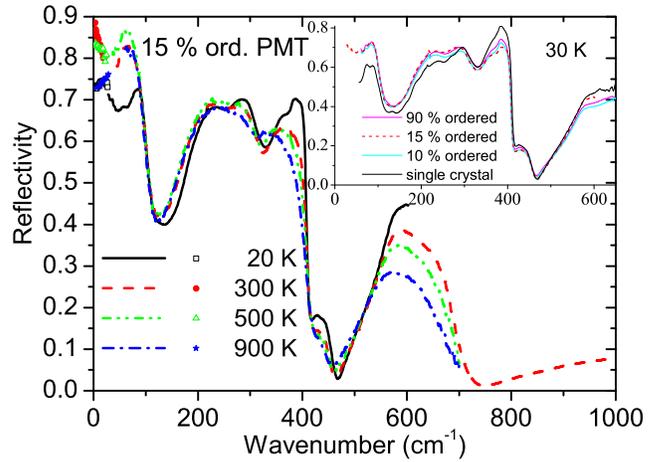}
  \end{center}
    \caption{(color online) Infrared reflectivity spectra of the 15\% ordered PMT ceramics at various temperatures.
    The low-frequency dots below 30\cm\, were calculated from the THz dielectric spectra (see
Eq.(~\ref{refl})). 30\K\, IR spectra of the single crystal and
different PMT ceramics
    are compared in the inset.}
    \label{Fig5}
\end{figure}

\begin{figure}
  \begin{center}
    \includegraphics[width=80mm]{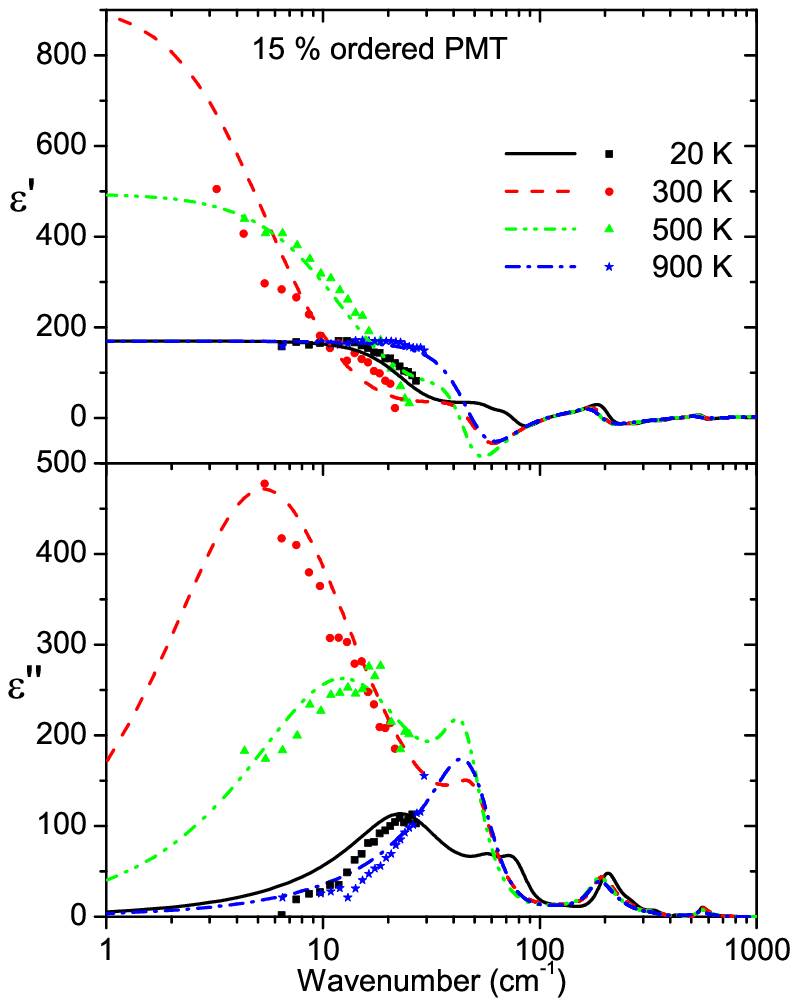}
  \end{center}
    \caption{(color online) Complex dielectric spectra of the 15\% ordered PMT ceramics obtained from the fit of IR and THz reflectivity.
Note the log frequency scale and experimental THz data (solid
dots).}
    \label{Fig6}
\end{figure}

$\varepsilon_{R\infty}$ in Eq.~\ref{integ1} marks a sum of phonon
and electron contributions into the static permittivity. Solution
of the integral equations (Eq.~\ref{integr}) is known as a
solution of an ill-posed problem and the most general method for
its solution is the Tikhonov regularization
\cite{tikhonov,groetsch}, which we used for calculation of the
distribution function of relaxation times $f$($\tau$) at different
temperatures. The results below 400\K\, are shown in
Fig.~\ref{Fig3}. The distribution function at higher temperatures
was not evaluated since the relaxation moves into the narrow THz
range, in the MW range no data are available and almost no
dispersion is observed below 1 GHz. Nevertheless, we will show
below that the THz relaxation above room temperature merges with
the lowest-frequency E symmetry phonon and disappears above
$T_{d}$.

One can see from Fig.~\ref{Fig3} that at 400\K\, only a narrow distribution $f$($\tau$)
appears in the MW range, but on cooling the relaxation slows down and broadens so that it
is stretched between kHz and THz range at 200\K. Temperature dependence of the longest
($\tau_{L}$), most probable ($\tau_{mp}$) and shortest ($\tau_{S}$) relaxation times are
plotted in Fig.~\ref{Fig4} together with their fits. $\tau_{S}$($T$) was fitted to the
Arrhenius law
\begin{equation}\label{Arrhenius}
\tau_{S}(T)=\tau_{S\infty}exp\left(\frac{E_{a}}{T}\right),
\end{equation}
with $\tau_{S\infty}$=7.9$\ast10^{-15}$\,s and $E_{a}$=988\K. $\tau_{mp}$ and $\tau_{L}$
follow the Vogel-Fulcher formula
\begin{equation}\label{vogel1}
\tau_{L,mp}(T)=\tau_{L,mp\infty}exp\left(\frac{E_{L,mp}}{T-T_{VF_{L,mp}}}\right),
\end{equation}
with $\tau_{mp\infty}$=1.6$\ast10^{-13}$\,s, $E_{mp}$=553\K, $T_{VF_{mp}}$=146\K\, and
$\tau_{L\infty}$=8.3$\ast10^{-13}$\,s, $E_{L}$=809\K\, and $T_{VF_{L}}$=152\K. One can
see that with increasing fitted relaxation time the activation energy and freezing
temperature increase. We have fitted as well the frequency of dielectric loss maxima
$\varepsilon''_{max}(T)$ by means of the Vogel-Fulcher formula which yields the freezing
temperature $T_{VF}$=140\K, $f_{\infty}$=1.73.10$^{13}$ Hz and $E$=875\K\, (see inset in
Fig.~\ref{Fig1}). It is in reasonable agreement with parameters of previous fits in
Fig.~\ref{Fig4}, but we stress that the relaxation times $\tau_{mp}$ and $\tau_{L}$ have
better physical meaning than $\varepsilon''(T)_{max}$.

We note that our freezing temperatures in 90\% ordered ceramics
are slightly ($\sim$20\K) higher than the published $T_{VF}$ for
disordered single crystal\cite{ko03}. This small difference is not
caused by ordering in the B sites since it has an opposite effect;
$T_{VF}$ usually decreases with rising ordering.\cite{wang05} The
higher $T_{VF}$ in the ordered ceramics is presumably caused by
the 5\% admixture of PbZrO$_{3}$ in PMT, as also observed in
Refs.\cite{akbas97,akbas01}

Note the splitting of the relaxation times distribution function $f(\tau)$ at 230\K,
better seen at 215\K\, (Fig~\ref{Fig3}). Similar splitting was observed also in other
relaxors\cite{macutkevic06,bovtun06}. Long-time relaxation was assigned to flipping and
short-time part of $f(\tau)$ to breathing of polar clusters. The former part disappears
from the spectra below freezing temperature, while the latter part anomalously broadens
on cooling. Frequency independent $\varepsilon^{\prime\prime}$ (i.e. almost uniform
$f$($\tau$)) spectra are seen below 150\K\, giving evidence about anomalously broad
distribution of relaxation frequencies at low temperatures. This behavior appears to be
general since it was observed in all lead-based perovskite relaxor ferroelectrics so far
investigated in a sufficiently broad frequency and temperature range.\cite{petzelt07}
This effect can be understood from the broad distribution of activation energies for
hopping of the Pb cations in random fields and bonds of chemically disordered cations
with two different valences in the perovskite B sites.\cite{rychetsky03}

The relaxation (i.e. central peak) was revealed also in the
inelastic neutron scattering (INS)
spectra.\cite{gvasalia03a,gvasalia04a} Its temperature behavior
was not investigated, but the neutron scattering experiment showed
that the diffuse scattering intensity and the amplitude of the
lead displacements exhibit similar temperature
dependences.\cite{gvasalia04b}

We investigated also the higher frequency region above 100\,GHz by means of THz and IR
spectroscopy at temperatures from 20 up to 900\K. The IR reflectivity above 1000\cm\,
exhibits almost no frequency dispersion, therefore the IR spectra in Fig.~\ref{Fig5} are
plotted only up to 1000\cm\,(30\,THz). The rich phonon structure was observed only below
700\cm. We measured the IR reflectivity of the single crystal as well as of variously
ordered PMT ceramics and the comparison of 30\K\, spectra in inset of Fig.~\ref{Fig5}
shows that there is appreciably no difference between the polar phonon structure of
variously ordered ceramics. Phonon frequencies are also the same for the (B-site
disordered) single crystal, only the mode strengths are slightly different.

In order to obtain all polar phonon parameters as a function of
temperature, IR and THz spectra were fitted simultaneously using
the generalized-oscillator model with the factorized form of the
complex permittivity:\cite{gervais83}
\begin{equation}\label{eps}
\varepsilon^{*}(\omega)=\varepsilon_{\infty}\prod_{j}\frac{\omega^{2}_{LOj}-\omega^{2}+i\omega\gamma_{LOj}}{\omega^{2}_{TOj}-\omega^{2}+i\omega\gamma_{TOj}}
\end{equation}
where $\omega_{TOj}$ and $\omega_{LOj}$ denotes the transverse and longitudinal frequency
of the j-th polar phonon, respectively, and $\gamma$$_{TOj}$ and $\gamma$$_{LOj}$ denotes
their corresponding damping constants. $\varepsilon$$^{*}$($\omega$) is related to the
reflectivity R($\omega$) by
\begin{equation}\label{refl}
R(\omega)=\left|\frac{\sqrt{\varepsilon^{*}(\omega)}-1}{\sqrt{\varepsilon^{*}(\omega)}+1}\right|^2
.\end{equation}

The high-frequency permittivity $\varepsilon_{\infty}$ resulting from the electronic
absorption processes was obtained from the room-temperature frequency-independent
reflectivity tails above the phonon frequencies (i.e. above 1000\cm) and was assumed to
be temperature independent.

The real and imaginary parts of $\varepsilon^*$($\omega$) obtained
from the fits to IR and THz spectra (including the THz
experimental points) are shown in Fig.~\ref{Fig6}. Parameters of
the fits performed at 20, 300 and 900\K\, are summarized in Table
I. We resolved 15 polar modes in the IR spectra at 20\K, 11 at
300\K\, and 8 modes at 900\K.

\begin{table*}
\caption{Parameters of polar phonon modes in 15\% ordered PMT ceramics obtained from the
fit to IR and THz spectra at 20, 300 and 900\,K. Frequencies $\omega_{TOi}$,
$\omega_{LOi}$ and dampings $\gamma_{TOi}$, $\gamma_{LOi}$ of modes are in
\ensuremath{\mbox{cm}^{-1}}, $\Delta \varepsilon_{i}$ is dimensionless,
$\varepsilon_{\infty}$=4.75. Parameters of the mode No. 15 were not determined at 20\K,
because the mode lies beyond the frequency window of our cryostat.}
\resizebox{17.4cm}{!}{
\begin{tabular}{|r | c c c c c||c c c c c ||c c c c c|}\hline
  &\multicolumn{5}{c||}{20 K}&\multicolumn{5}{c|}{300 K}&\multicolumn{5}{c|}{900 K}\\
  \hline
  No&$\hspace{0.2cm} \omega_{TOi} \hspace{0.2cm}$&\hspace{0.2cm}
  $\gamma_{TOi}$ \hspace{0.2cm}&\hspace{0.2cm} $\omega_{LOi}$ \hspace{0.2cm}&\hspace{0.2cm}
  $\gamma_{LOi}$ \hspace{0.2cm}&\hspace{0.2cm} $\Delta
  \varepsilon_{i}$ \hspace{0.2cm}&
  $\hspace{0.2cm} \omega_{TOi} \hspace{0.2cm}$&\hspace{0.2cm}
  $\gamma_{TOi}$ \hspace{0.2cm}&\hspace{0.2cm} $\omega_{LOi}$ \hspace{0.2cm}&\hspace{0.2cm}
  $\gamma_{LOi}$ \hspace{0.2cm}&\hspace{0.2cm} $\Delta
  \varepsilon_{i}$ \hspace{0.2cm}&
   $\hspace{0.2cm} \omega_{TOi} \hspace{0.2cm}$&\hspace{0.2cm}
  $\gamma_{TOi}$ \hspace{0.2cm}&\hspace{0.2cm} $\omega_{LOi}$ \hspace{0.2cm}&\hspace{0.2cm}
  $\gamma_{LOi}$ \hspace{0.2cm}&\hspace{0.2cm} $\Delta
  \varepsilon_{i}$ \hspace{0.2cm}
 \\ \hline
 1&25.5&30.8&41.8&42.4&115.2&15.4&49.3&38.6&42.9&835.6&&&&&\\
 2&61.0&17.6&62.4&16.8&3.4&&&&&&&&&&\\
 3&79.8&36.4&107.3&29.2&24.6&52.4&31.9&106.9&39.0&60.2&52.3&43.4&104.6&34.7&136.9\\
 4&136.3&96.6&145.1&56.3&3.4&143.7&138.3&146.1&88.6&1.1&146.5&235.4&148.4&112.6&1.5\\
 5&203.3&59.3&261.8&60.9&15.7&192.3&64.8&248.6&40.8&18.8&185.4&65.3&309.0&50.8&24.8\\
 6&267.6&52.8&329.8&62.0&0.87&250.0&41.4&273.6&50.0&0.30&&&&&\\
 7&283.0&10.5&329.6&61.1&0.01&274.7&47.1&329.8&56.1&0.14&&&&&\\
 8&343.3&48.7&373.8&70.5&0.44&339.4&26.8&341.6&30.0&0.19&312.1&53.7&411.3&31.3&0.23\\
 9&386.3&30.3&388.5&27.4&0.06&344.3&57.8&412.1&20.6&0.32&&&&&\\
10&400.9&76.0&409.2&11.9&0.43&&&&&&&&&&\\
11&411.8&48.4&433.0&23.5&0.15&&&&&&&&&&\\
12&434.8&26.2&458.8&27.8&0.03&424.9&39.7&451.5&37.3&0.11&426.1&47.2&445.2&46.5&0.10\\
13&515.6&56.6&522.0&55.4&0.18&&&&&&&&&&\\
14&559.6&56.0&700.0&77.0&1.04&560.5&65.0&642.9&117.8&1.07&535.9&112.0&646.5&104.6&1.19\\
15&?&?&?&?&?&649.3&85.9&700.0&77.4&0.04&648.4&90.4&690.5&79.0&0.01\\
  \hline
  \hline
\end{tabular}
}
\label{IRmodes}
\end{table*}

Let us compare the number of observed polar modes with the
prediction of the factor group analysis (see Table II). We remind
that the structural study of PMT down to 1.5\K\, showed no
macroscopic phase transition, the structure remaining cubic at all
temperatures.\cite{gvasalia04b} Simple cubic perovskites of space
group $Pm\bar{3}m$ have 3$F_{1u}$ IR-active polar modes, but PMT
has two kinds of B-site cations which are 1:1 short range
ordered,\cite{akbas97} therefore the local space group is rather
$Fm\bar{3}m$ with 4$F_{1u}$ IR-active phonons.\cite{hlinka06} This
is, however, appreciably less than the observed number of polar
phonons in Table I. Since the IR spectra are sensitive to local
symmetry already in the range of a few unit cells and because of
the presence of polar nanoclusters in the lattice, the IR
selection rules should allow more modes below $T_d$. If we assume
that the local symmetry of the polar nanoclusters is rhombohedral
$R3m$ with 2 formula units per unit cell (like in PMN-PT), then 16
modes are simultaneously active in the IR and Raman spectra below
$T_d$ (see Table II). We have observed 15 modes at 20\K, which
reasonably agrees with the expectation. The fact that only 11
modes are seen at 300\K, can be explained by higher damping of the
modes at RT, which may cause overlapping of some modes.

The situation at high temperatures requires additional discussion. No polar clusters are
expected above $T_d$ so that only 4 IR modes are permitted for the space group
$Fm\bar{3}m$, but we resolved 8 polar modes in the IR spectra up to 900\K. Factor group
analysis for the $Fm\bar{3}m$ structure assumes 1:1 chemical ordering in the B
sites\cite{hlinka06} in small chemical clusters (3-150 nm in PMT)\cite{akbas01,wang05},
while the rest of the sample volume (e.g. 90\% in the 10\% ordered sample) remains
chemically disordered, i.e. the Mg$^{2+}$ and Ta$^{5+}$ cations are randomly dispersed
over both $\beta'$ and $\beta''$ sites. This could have some influence on the IR
selection rules. Moreover, even in the chemically ordered nanoregions the $\beta''$
sublattice is not occupied by one cation, but it is randomly occupied by Ta and Mg
cations in the 1:2 ratio. It is natural to expect that all lattice vibrations where the B
cations are involved, could be split due to the chemical inhomogeneity in the $\beta''$
sublattice. The reason for the mode splitting in chemically disordered solids is
discussed in detail in the review by Barker and Sievers.\cite{barker75} We may expect
that the modes involving Pb vibration (1$F_{1u}$+1$F_{2g}$) do not split, while the rest
of the modes (including oxygen vibrations against $\beta'$ and $\beta''$) from the second
column in Table II split into doublets. The new factor group analysis of
Pb($\beta'_{1/2}$$\beta''_{1/2}$)O$_{3}$ above $T_{d}$, where $\beta'$=Ta and
$\beta''$=Ta$_{1/3}$Mg$_{2/3}$ is shown in Table II. This analysis predicts 8 IR modes in
full agreement with the results in Table I: The TO2, TO3 (200-400\cm) and TO4 modes
(500-700\cm) are split because these so called Slater and Axe modes\cite{hlinka06}
involve the B-cation vibrations. On the other hand, the TO1 mode near 90\cm\, (so called
Last mode) is not split above $T_d$, because it involves mostly translation of Pb cations
against the O octahedra.

Let us mention the result of the INS experiment on PMN single
crystal by Vakhrushev and Shapiro.\cite{vakhrushev02} They
observed splitting of the TO1 mode also far above $T_{d}$ and
explained it by a two mode behavior due to the chemical disorder.
However, their results were not confirmed by other groups
performing INS experiments on relaxors.

Our PMT ceramics contain 5\% of Zr and the structural studies have
shown,\cite{dmowski02,cantoni04} that the $\beta'$ sublattice is
occupied only by Ta cations, while all Mg, Zr and rest of Ta
cations occupy the $\beta''$ sublattice. Heterogeneous occupation
of the $\beta''$ sublattice with three cations of strongly
different mass could cause even more complicated splitting of the
modes than that observed (see Table II).
\begin{table*}
\caption{Symmetry classification and activity of optical
vibrational modes for different hypothetical local PMT structures.
IR and R mean activity in IR and Raman spectra, respectively. (-)
marks silent mode. Sublattice $\beta'$ marks Ta and sublattice
$\beta''$ is occupied by Mg$_{2/3}$Ta$_{1/3}$ in the pure 1:1
ordered PMT\cite{akbas97}.}

\begin{tabular}{|l l l l|}\hline \hline
ABO$_{3}$ &A(B'$_{1/2}$B''$_{1/2}$)O$_{3}$ &A($\beta'_{1/2}$$\beta''_{1/2}$)O$_{3}$
&A(B'$_{1/2}$B''$_{1/2}$)O$_{3}$
\\
$Pm\bar{3}m$    &$Fm\bar{3}m$   &$Fm\bar{3}m$   &$R3m$\\
Z=1 &Z=2 &Z=2 &Z=2\\ \hline
3$F_{1u}$(IR) &3$F_{1u}$(IR) &6$F_{1u}$(IR) &3$A_{1}$(IR,R)\\
&&&3$E$(IR,R)\\
1$F_{2u}$(-) &1$F_{2u}$(-) &2$F_{2u}$(-)&1$A_{2}$(-)\\
             &             & &1$E$(IR,R)\\
progenitors &1$F_{1u}$(IR)&2$F_{1u}$(IR)&1$A_{1}$(IR,R)\\
are \textbf{q}$\neq$0& & &1$E$(IR,R)\\
modes &1$A_{1g}$(R)&2$A_{1g}$(R) &1$A_{1}$(IR,R)\\
 &1$E_{g}$(R)&2$E_{g}$(R) &1$E$(IR,R)\\
    &2$F_{2g}$(R)&3$F_{2g}$(R)  &2$A_{1}$(IR,R)\\
    &   &   &2$E$(IR,R)\\
    &1$F_{1g}$(-)&2$F_{1g}$(-)  &1$A_{2}$(-)\\
    &1$F_{2u}$(-)&2$F_{2u}$(-)  &1$A_{2}$(-)\\
    &   &   &1$E$(IR,R)\\ \hline
3$F_{1u}$+11$F_{2u}$ &$A_{1g}$+$E_{g}$+$F_{1g}$&2$A_{1g}$+2$E_{g}$+2$F_{1g}$ &7$A_{1}$+2$A_{2}$+9$E$\\
    &4$F_{1u}$+2$F_{2g}$+$F_{2u}$&8$F_{1u}$+3$F_{2g}$+2$F_{2u}$  &\\ \hline
3 IR + O R &4 IR + 4 R&8 IR + 7 R &16(IR+R)\\ \hline
  \hline
\end{tabular}
\label{IRmodes}
\end{table*}

PMT is isostructural with PMN, therefore IR spectra are expected similar for both
compounds. However, only 8 modes were resolved in the PMN spectra at RT and 20\K\, and
also the characteristic deep minimum seen in PMT near 300\cm\, was not resolved in the
PMN spectra.\cite{prosandeev04,hlinka06b} The difference may be understood by a larger
chemical disorder and higher anharmonicity of Nb vibrations in PMN, which causes a larger
effective phonon damping and therefore possible overlapping of modes. Remarkable
influence of the B-site ordering on the phonon damping was observed also in
PbSc$_{1/2}$Ta$_{1/2}$O$_{3}$ (PST).\cite{kamba05b} Nevertheless, it is worth to note
that our preliminary high-temperature IR reflectivity spectra of PMN as well as recently
published IR spectra of PLZT\cite{buixaderas07} obtained above $T_{d}$, reveal 6 polar
modes, which also cannot be explained assuming the simple ordering. Similar explanation
as presented here for PMT, may be, therefore, considered.

\begin{figure}
  \begin{center}
    \includegraphics[width=80mm]{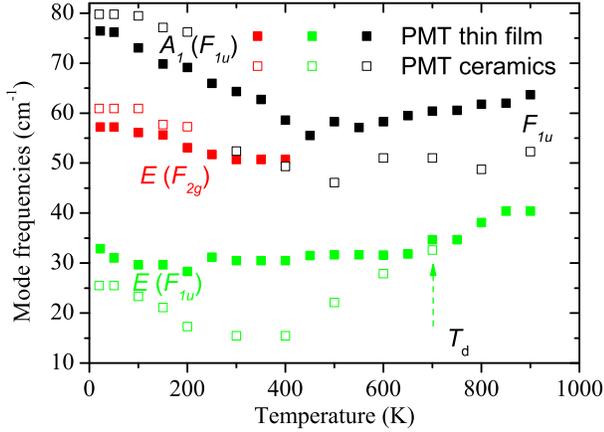}
  \end{center}
    \caption{(color online) Temperature dependences of the three lowest-frequency IR modes in 15\% ordered
    PMT ceramics (open points)
    and disordered thin films (solid points). Symmetry of the modes is marked, in parenthesis the symmetry of the
    modes above Burns temperature $T_{d}$ is shown.}
    \label{Fig7}
\end{figure}

\begin{figure}
  \begin{center}
    \includegraphics[width=70mm]{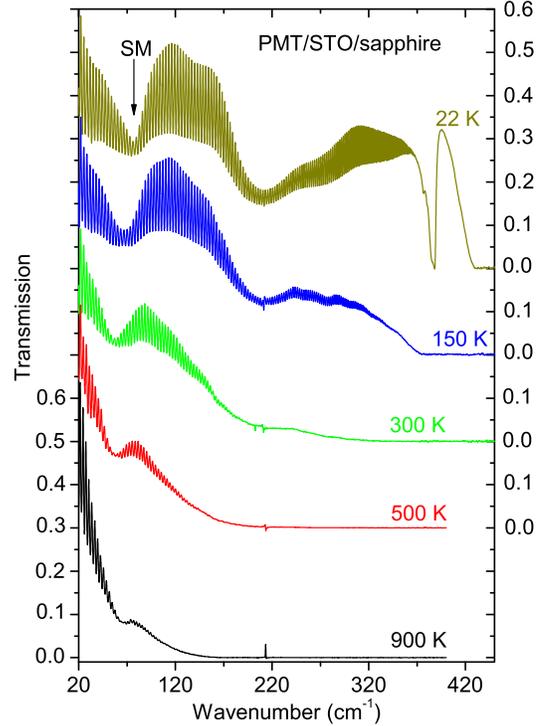}
  \end{center}
    \caption{(color online) Temperature dependence of the IR transmission spectra of PMT thin film (500\,nm) on
    sapphire substrate (500\,$\mu$m) with STO buffer layer (40\,nm). Note the change of scale for each spectrum.}
    \label{Fig8}
\end{figure}

An interesting temperature dependence of the mode frequencies was
observed below 100\cm\, (see Fig.~\ref{Fig7}). Far-IR (FIR)
spectra are less accurate, because FIR spectral range is
experimentally difficult attainable. Therefore we investigated
both FIR reflectivity of ceramics and FIR transmission spectra of
thin films (sapphire substrate is transparent only in the FIR
range).

FIR transmission spectra of PMT thin film were measured at 20 - 900\K\, and shown at
selected temperatures in Fig.~\ref{Fig8}. Dense and large-amplitude oscillations in the
spectra are due to the multi-pass interferences in the substrate, while the broad minima
correspond to the frequencies of polar modes in the PMT film. The transmission decreases
on heating mainly due to the increase in the absorption of the sapphire substrate.

The spectra of a bare sapphire substrate, STO buffer thin film on
sapphire and the PMT film on the substrate with STO buffer layer
were determined independently for each temperature studied.
Temperature dependence of the IR transmission spectra of sapphire
substrate was published in Ref.\cite{kamba05}, the spectra of STO
buffer layer will be discussed elsewhere.\cite{nuzhnyy} For a
given temperature, the transmission spectrum of the sapphire
substrate was fitted with a sum of classical damped oscillator
model for the dielectric function (using the Fresnel formulae for
coherent transmission of a plane-parallel sample, i.e. taking into
account the interference effects\cite{Born})
\begin{equation}
\label{eps3p}
 \varepsilon^*(\omega)
 = \varepsilon_{\infty} + \sum_{j=1}^{n}
\frac{\Delta\varepsilon_{j}\omega_{TOj}^{2}} {\omega_{TOj}^{2} -
\omega^2+\textrm{i}\omega\gamma_{TOj}}
\end{equation}
where $\Delta\varepsilon_{j}$ means the contribution of the j-th mode to the static
permittivity. The rest of the parameters in Eq.~\ref{eps3p} has the same meaning as in
Eq.~\ref{eps}.

The resulting parameters of the phonons in sapphire substrate were used for the fit of
the STO/sapphire and PMT/STO/sapphire two- and three-layer systems, respectively. The
complex transmittance was computed by the transfer matrix formalism method including
interference effects.\cite{Heavens}

Let us compare the temperature dependence of the low-frequency
polar modes obtained from the IR reflectivity and THz transmission
spectra of ceramics with the modes calculated from the IR
transmission spectra of the thin film (see Fig.~\ref{Fig7}). In
ceramics one can see (follow the open points) an unsplit TO1 mode
above $T_{d}\sim$ 700\K. The mode splits into two components below
$T_{d}$ confirming the appearance of polar clusters with lower
local symmetry. The higher frequency TO1 component (of $A_{1}$
symmetry) slightly softens on cooling to 500\K, but on further
cooling it pronouncedly hardens, indicating that we see the local
ferroelectric soft mode inside the polar clusters.

The second ($E$) component of the TO1 mode exhibits softening below $T_{d}$ and hardening
below 300\K. A new mode appears near 58\cm\, below $T_{max}\sim$200K. This mode stems
from the Raman active mode\cite{lushnikov04} of $F_{2g}$ symmetry above $T_{d}$ and it
has $E$ symmetry at low temperatures. It becomes resolved at low temperatures due to the
breaking of local symmetry as well as due to the lower phonon damping of both TO1
components. Similar temperature behavior of all three phonons was observed in PMN, PST,
PLZT and also in other relaxor solid solutions\cite{kamba05,kamba05b,macutkevic06,
buixaderas07} and it seems that it is a general behavior in all lead-based perovskite
relaxors.

Let us mention that we observed no overdamping of TO1 phonon below $T_{d}$ frequently
seen in INS spectra of other relaxors, which is called phonon "waterfall
effect".\cite{hirota06} TO1 phonon is underdamped at all temperatures indicating that the
"waterfall effect" in the INS spectra is just an apparent effect due to a coupling of
heavily damped soft optic mode with some other excitation, e.g. acoustic phonon
branches.\cite{hlinka03}

Gvasaliya $\it et\, al.$\cite{gvasalia03b} observed in INS spectra (but did not explain)
a remarkable change of the vibrational state density in the range of 50--80\cm\, on
cooling from 300 to 50\K. This change could be connected with the hardening of the optic
$A_{1}$ modes on cooling seen in Fig.~\ref{Fig7}.

IR transmission spectra of the thin film yield slightly different mode frequencies (see
solid dots in Fig.~\ref{Fig7}) and moreover, the TO1 mode is split at all temperatures up
to 900\K. Different mode frequencies as well as different temperatures of the TO1
splitting in ceramics and thin films can be caused by the strain which induces anisotropy
in the film. Nevertheless, we observed that the film is rather relaxed, strain is smaller
than 0.1\%, therefore the strain effect on the phonon spectra should be small. The film
is mostly (111) oriented and along the same direction is oriented also the spontaneous
polarization in rhombohedral phase of PMT. If we assume that the local symmetry in polar
clusters is also rhombohedral and that the $E$ modes are IR active for polarization of IR
radiation perpendicular to (111), i.e. in the plane of the thin films, than it is clear
that the intensity of $E$ modes should be enhanced in IR spectra of our thin film. This
is probably reason, why the $E$ modes are better resolved in high temperature IR spectra
of the film, while their intensity remarkably weaken in reflectivity spectra of ceramics
on heating.

We note that the accuracy of the phonon frequency determination is different in various
experimental techniques used. IR reflectivity of ceramics above 50\cm\, is less accurate
than the FIR transmission spectra of thin films. Therefore the mode frequencies above
50\cm\, in thin film are determined more accurately. This is probably also the reason why
the $E$($F_{2g}$) mode is resolved in IR transmission spectra already at 400\K, i.e. by
200\K\, higher than in the less sensitive IR reflectivity spectra. On other hand, the THz
transmission spectra of the ceramics below 30\cm\, are more accurate than the FIR
transmission spectra of the thin films, therefore we believe more to the lowest frequency
$E$ mode obtained from the THz spectra of ceramic samples (especially below 400\K).
Simultaneously it is probable that the $E$ mode is stiffened in the thin film due to
small strain, as frequently observed e.g. in STO thin films.\cite{petzelt07}

Second reason for different frequency of the $E$ mode in ceramics and thin films can be
also different strength of relaxation mode in ceramics and films. The relaxation mode
expresses the dynamics of polar clusters, i.e. their flipping and breathing. It appears
in THz dielectric spectra below $T_{d}$ together with the $E$ mode and both modes are
overlapped because they have similar frequencies. The relaxation mode slows down to
microwave and broadens on cooling. It is partly seen in the THz dielectric spectra in
Figs.~\ref{Fig2} and ~\ref{Fig6} and it is manifested by the highest permittivity and
losses near 1\,THz close below $T_{d}$ (not shown in the figures). Below 150\K\, the
relaxation does not contribute to the THz spectra anymore so that the $E$ symmetry phonon
mode is well resolved at low temperatures (e.g. at 20\K\, spectra in Fig.~\ref{Fig6}).
Permittivity in the film is lower than in the ceramics. It means that the dielectric
strength of the relaxation is higher in ceramics than in the film and the relaxation
frequency is lower in the ceramics than in the film. THz and IR spectra were fit below
40\cm\, with one effective mode, but because of different relaxation frequency also the
peak in $\varepsilon''(\omega)$ seen as $E$ mode in Fig.~\ref{Fig7} can be different in
both kinds of PMT samples.

Nevertheless, it seems that the Burns temperature (i.e. splitting temperature of the TO1
mode) is different in ceramics ($\sim$ 900\K) and in the thin PMT film (above 900\K). In
disordered PMT single crystals, other authors\cite{ko03,korshunov92} found $T_{d}$ close
below 600\K. The low-frequency dielectric studies of PMT ceramics show $T_{d}$ even only
430\K\, (estimated from the linear fit of 1/$\varepsilon(T)$), which was independent of
the degree of chemical order.\cite{wang05} Nevertheless, note that these dielectric
measurements were performed only below 500\K, i.e far below the expected $T_{d}$.
Therefore we claim that different values of $T_{d}$ published by various authors can be
caused both by different sensitivity of various experimental techniques as well as by
different degree of chemical order. Further experiments are necessary for distinguishing,
which factor is more important.

\section{Conclusion}

Broad-band dielectric spectra from 100 Hz to 90 THz revealed a dielectric relaxation
which appears below the Burns temperature $T_{d}$ close to the soft phonon mode frequency
near 1 THz. The relaxation expresses dynamics of polar clusters and it slows down on
cooling from the THz region through MW and radio-frequency region and simultaneously
broadens anomalously giving rise to frequency independent losses at low temperatures. The
relaxation frequency is only slightly influenced by the degree of chemical order in the
perovskite B sites. In ceramic samples the TO1 phonon shows splitting below $T_{d}$ due
to the local breaking of the cubic symmetry in polar clusters. In the thin film the TO1
phonon is split at all investigated temperatures below 900\K. Both $A_{1}$ and $E$
components of the TO1 mode remarkably harden below 300-400\K\, indicating their
assignment to ferroelectric soft modes inside the polar clusters. Twice higher number of
polar modes than expected from the factor group analysis for the $Fm\bar{3}m$ cubic
structure was observed above $T_{d}$. This discrepancy was accounted for by the chemical
disorder in the perovskite B sites. On the other hand, we have shown that the degree of
chemical order has insignificant influence on the phonon frequencies. Slightly different
phonon frequencies in chemically disordered PMT thin film (in comparison to ceramics) can
be explained by different relaxation frequency which overlaps the $E$ mode as well as by
different spectral techniques used in the study of the thin film and bulk samples.

Finally we conclude that the observed behavior of the dielectric
relaxation as well as of the soft phonons appears the same for all
lead-based relaxor ferroelectrics with the perovskite structure so
far studied.

\begin{acknowledgments}
The work was supported by the Grant Agency of the Czech Republic (Project No.
202/06/0403), AVOZ10100520 and Ministery of Education (OC101-COST539). We are grateful to
A. Kania for providing us with the PMT single crystal.
\end{acknowledgments}


\end{document}